# Unconventional Surface State Pairs in a High-Symmetry Lattice with Anti-ferromagnetic Band-folding

Lin-Lin Wang[1*], Junyeong Ahn[2], Robert-Jan Slager[3], Yevhen Kushnirenko[1], Benjamin G. Ueland[1], Aashish Sapkota[1], Benjamin Schrunk[1], Brinda Kuthanazhi[4], Robert J. McQueeney[1,4], Paul C. Canfield[1,4] and Adam Kaminski[1,4]

[1]Ames National Laboratory, U.S. Department of Energy, Ames, IA 50011, USA
[2]Department of Physics, Harvard University, Cambridge, MA 02138, USA
[3]TCM Group, Cavendish Laboratory, University of Cambridge, J. J. Thomson Avenue, Cambridge CB3 0HE, UK
[4]Department of Physics, Iowa State University, Ames, IA 50011, USA

*llw@amelsab.gov

## Abstract

Many complex magnetic structures in a high-symmetry lattice can arise from a superposition of well-defined magnetic wave vectors. These "multi-q" structures have garnered much attention because of interesting real-space spin textures such as skyrmions. However, the role multi-q structures play in the topology of electronic bands in momentum space has remained rather elusive. Here we show that the type-I anti-ferromagnetic 1q, 2q and 3q structures in an face-centered cubic sublattice with band inversion, such as NdBi, can induce unconventional surface state pairs inside the band-folding hybridization bulk gap. Our density functional theory calculations match well with the recent experimental observation of unconventional surface states with hole Fermi arc-like features and electron pockets below the Neel temperature. We further show that these multi-q structures have Dirac and Weyl nodes. Our work reveals the special role that band-folding from anti-ferromagnetism and multi-q structures can play in developing new types of surface states.

# I. Introduction

Rare-earth monopnictides (RPn) in the rock-salt crystal structure have been studied as promising materials for spintronics[1]. These compounds have attracted renewed interests because of emerging topological properties. The non-magnetic LaPn were first predicted as topologically non-trivial due to the band inversion between La $d$ and Pn $p$ orbitals[2]. The crossover and trend of such $d$-$p$ band inversion in the RPn series have been studied by experiments[3-8] and calculations[9, 10], which have largely focused on the paramagnetic state. The gradual filling of the $4f$ orbitals and change of spin-orbit coupling (SOC) strength across the RPn series can potentially bring high tunability in the multitude of topological states when coupled with magnetic properties[11-16]. Although some RPn compounds, such as CeSb[17], GdBi[18], and CeBi[19], have been proposed to host magnetic Weyl points (WPs) in the ferromagnetic (FM) state when fully magnetized, most RPn prefer anti-ferromagnetic (AFM) configurations. Angle-resolved photo-electron spectroscopy (ARPES) study of the RPn in the AFM state is rare except for CeBi[20, 21].

Very recently ARPES[22] has discovered unconventional surface states (SS) in NdBi below the Neel temperature ($T_N$=24 K) that display a combination of hole Fermi arc-like features and SS electron pockets, as well as a peculiar temperature-dependent splitting. Unlike the Fermi arcs in cuperates[23] induced by a pseudo-gap opening and those in Weyl semimetals[24] by the opposite chirality of WPs, the origin of the newly discovered hole Fermi arc-like features in AFM NdBi is unclear. Figures 1(a) and (b) show the ARPES data on NdBi (001) above and below $T_N$, respectively. Above $T_N$ and similar to other paramagnetic RBi, a large 4-fold star-shaped bulk hole pocket and another small one at the $\bar{\Gamma}$ point overlap with the electron pocket projected from the $X$ point in Fig.1(a). These results match reasonably well with the 2D surface Fermi surface (SFS) calculated with density functional theory[25, 26] (DFT) on the non-magnetic (NM) NdBi (001) in Fig.1(c). Surprisingly, when cooling below $T_N$ for NdBi (Fig.1(b)), ARPES observed 4-fold unconventional hole Fermi arc-like features and SS electron pockets emerging out of the bulk band projections. The hole Fermi arc-like features connect to the bulk hole pockets and the elliptical SS electron pockets have a sharp cusp. However, the DFT-calculated SFS on NdBi (001) with the 4-fold rotational symmetry in Fig.1(e) with the reported layer-stacking AFM structure[27] (referred as 1q) lacks these 4-fold unconventional SS features.

NdBi has a high-symmetry crystal structure without Nd-*4f*-level hybridization (e.g. Kondo effect), which makes the ARPES discovery more intriguing.

The related monopnictide compounds with actinides (5*f*) in the same rock-salt structure, such as UP, UAs, USb[28] and NpBi[29], have been reported before in type-I AFM multi-q structures, which are difficult to distinguish from the simple 1q using standard neutron diffraction techniques. The multi-q structures focused here are the commensurate magnetic structures on the Nd face-centered cubic (FCC) sublattice with superpositions of propagating vectors along the equivalent {001} directions, i.e., the type-I AFM multi-q structures[30-32]. In contrast, the type-II and III AFM on an FCC lattice have propagating vectors along the {111} or {½ 01} directions, respectively, giving different multi-q structures[30-32]. The magnetic moments in these multi-q structures can be noncollinear and noncoplanar. In general, multi-q has also been used to define the spin texture as superpositions of multiple spin helices or spin density waves, mostly incommensurate, such as skyrmions[33-42] in 2D thin films or layered structures, and hedgehog[43, 44] in 3D structures, with a real space non-trivial topology and also a periodicity much larger than the underlying crystal lattice. Although magnetic space groups (MSG) give all the symmetry operations for Wyckoff positions with breaking time-reversal symmetry due to magnetism, multi-q as a superposition of multiple propagating vectors can include the orientation of the magnetic moments to describe both commensurate and incommensurate magnetic structures with noncollinear and noncoplanar components.

Here, we explore the effects of the type-I AFM multi-q magnetic structures on the band structure and topology of NdBi using DFT calculation. We find that NdBi AFM multi-q structures with two (2q) and three (3q) wave vectors give rise to the 4-fold unconventional surface state pairs (SSP) on (001) surface as shown in Fig.1 (g)-(h) for 2q and Fig.1(j)-(k) for 3q, respectively; while 2q (010) and 1q (110) (equivalent to cubic (010)) have 2-fold SSP as shown in Fig.1(f) and (i), respectively. Considering the possibility of multiple domains of {001} directions in experimental samples, these unconventional SSP match well with the ARPES data below $T_N$. The hybridization gap from AFM band-folding along the {001} directions is the common theme in these multi-q structures of NdBi. The difference is the number of {001} directions that the AFM band-folding occurs. It occurs along only one {001} direction in 1q, but two and three equivalent {001} directions in 2q and 3q, respectively, giving the unconventional SSP of only 2-fold in 1q, 2-fold or 4-fold in 2q, and 4-fold in 3q. Additionally, we find that NdBi 1q and 2q are Dirac semimetals, and 3q is a Weyl semimetal. Our result shows that a Fermi arc-like signature on a surface does not always imply the presence of Weyl fermions in the bulk, as in the case of 1q and 2q, making an unambiguous discovery of magnetic Weyl semimetals more elusive.

## II. Results and Discussion

### II-A. Hybridization gap opening in multi-q bulk band structures

The three different type-I AFM multi-q structures of NdBi are shown in Fig.1 (d), (g) and (j) with the corresponding bulk band structures presented in Fig.2 (a), (c) and (d), respectively. The high symmetry *k*-points and Brillouin zones (BZ) are in Fig.2(b). The NM NdBi in space group (SG) of *Fm-3m* (or 225) is known to have a band inversion between the Bi *p* and Nd *d* orbitals with anti-crossing along the *Γ-X* direction to host a strong topological insulator (TI) state[2] (see Supplementary Note 1 and Fig.S1). The 1q is in MSG of $P_I4/mnc$ (or 128.410) and its band structure along the *Γ-M'* direction (Fig.2(a)) retains the same band inversion in NM (neck-narrowing region), while along the AFM stacking [001] direction, the band folding opens up a hybridization gap along the *Γ-Z* direction with a Dirac point (DP) at about 0.5 eV above the Fermi energy ($E_F$). The hybridization gap opening from the AFM band-folding of the band-inversion region is best seen in Fig.2(c) for 2q in MSG of $P_C4_2/nnm$ (or 134.481). The band-folding in 2q happens at the *X* point (half-way along the *Γ-M'* or [010] direction highlighted by the green arrows), where a bulk band gap of about 0.2 eV opens near $E_F$ due to the hybridization between the folded bands having the same Bi *p* orbital character. The original band inversion now becomes the upper valence bands and is pushed to higher energy. Interestingly, with the magnetic moments changing from [001] in 1q to in-plane directions in 2q, the hybridization gap along the *Γ-Z* direction is reduced and two new band crossings emerge to form two DPs. For the 3q in MSG of *Pn-3m'* (or 224.113), comparing its band structure in Fig.2(d) to 2q in Fig.2(c), the *Γ-X* and *Γ-Z* directions become equivalent and so are the hybridization gaps. With lifting of the band double degeneracy along the other directions in 3q, new band crossings emerge around the *Γ* point, giving WPs along the *Γ-R* direction, which will be discussed later together with the DPs in 2q because they are near or below the $E_F$.

With the inversion symmetry intact for all three magnetic structures, parity eigenvalues at time-reversal invariant momentum (TRIM) can be used to calculate $Z_4 = \left(\frac{1}{2}\right)\sum_{k \in TRIM}(n_k^+ - n_k^-)\ mod\ 4$, where $n_k^+$ and $n_k^-$ are the number of filled states with even and odd parity, respectively. Interestingly, $Z_4$=2 for all the three gapless phases in the different multi-q structures, because the sum of the difference in the number of opposite parity eigenvalues over TRIM does not change. NdBi 1q and 2q are Dirac semimetals, while 3q is a Weyl semimetal. Counting both $-k_z$ and $+k_z$ directions, the one pair of DPs in 1q are about 0.5 eV above the $E_F$, while the two pairs of DPs in 2q are near the $E_F$. The more detailed symmetry description of the multi-q band structures and the corresponding propagating vectors can be found in Supplementary Note 2.

**II-B. Unconventional surface state pairs**

Next, we discuss the surface states generated by the different type-I AFM multi-q structures. The surface spectral function of 2q (001) along the $\bar{\Gamma}$-$\bar{X}$-$\bar{\Gamma}$ direction is presented in Fig.3(b) and compared side-by-side to the equivalent NM (001) $\bar{\Gamma}$-$\bar{X}$-$\overline{M'}$ direction in Fig.3(a). From the bulk band projection on NM (001), the band inversion is near the $\bar{X}$ point with a dashed line, the middle point between the $\bar{\Gamma}$ and $\overline{M'}$ points. The two surface Dirac cones for this strong TI are overlapped at the $\overline{M'}$ point with two SS stemming from band inversion region toward the $\overline{M'}$ point. In echo with the folding of the bulk bands in Fig.2(c) for 2q, these SS of the overlapped surface Dirac cones are folded from the $\overline{M'}$ to $\bar{\Gamma}$ point in Fig.3(b). Simultaneously, a bulk hybridization gap is opened in 2q with two new unconventional SSP appearing in the hybridization gap along the $\bar{\Gamma}$-$\bar{X}$ direction from $E_F$– 0.150 eV to $E_F$. The original band inversion region is pushed up into higher valence bands as marked by the green dots.

The surface spectral functions on 3q (001) are plotted in Fig.3(c), 2q (001) in (d), 2q (010) in (e), and 1q (110) in (f), respectively. For all the four multi-q surfaces in Fig.3 with the AFM band-folding hybridization gap along the different {001} directions, a pair of unconventional SSP emerges inside the bulk hybridization gap with similar dispersion along the $\bar{\Gamma}$-$\bar{X}$ direction (or $\bar{\Gamma}$-$\bar{Z}$ for 1q (110)). The SSP starts from the overlapped bulk band projection near the $\bar{\Gamma}$ point and the splitting increases toward the surface Brillouin zone (SBZ) boundary. Then along the $\bar{X}$-$\bar{M}$ (3q and 2q (001)), $\bar{X}$-$\bar{R}$ (2q (010)) or $\bar{Z}$-$\bar{R}$ (1q (110)) directions, the upper SS quickly merges with the bulk electron pocket, while the lower SS merges with the bulk hole pocket. Thus, the upper SS is of electron character, while the lower SS is of hole character, respectively. In contrast, on 1q (001) and (010) (or cubic (110)) surfaces, there are no such unconventional SSP (see Supplementary Note 3 and Fig.S2).

The SFS are plotted at $E_F$–0.100 eV for 3q (001) (Fig.3(k)) and $E_F$–0.050 eV for 2q (001) (Fig.3(l)). The hole SS merges with the bulk hole pocket when going off the $\bar{\Gamma}$-$\bar{X}$ direction, showing a Fermi arc-like feature. The electron SS forms an elliptical pocket by itself with a cusp toward the $\bar{X}$ point. Both of these SSP features are in good agreement with ARPES[22] near the $\bar{\Gamma}$ point. The electron SS pocket comes from the upward dispersion of the upper SS near the $\bar{X}$ point (Fig.3 (c)-(d)). At $E_F$, the SFS in Fig.3 (g)-(h) clearly show the upper SS envelops the bulk electron pocket when going off the $\bar{\Gamma}$-$\bar{X}$ direction. For 2q bulk band structure, the hybridization gap only appears along the $\Gamma$-$X$, not the $\Gamma$-$Z$ direction as seen in Fig.2(c). As a result, in the SFS on 2q (010) (Fig.3(i) and (m)), the SSP hole Fermi arc-like feature and electron pocket also only appear along the $\bar{\Gamma}$-$\bar{X}$ direction, while along the $\bar{\Gamma}$-$\bar{Z}$ direction is mainly bulk band projection, giving a clearly 2-fold symmetry. Similarly, on 1q (110) surface (Fig.3(j) and (n)), the AFM band-folding hybridization gap only appears along the $\Gamma$-$Z$ direction and the SSP is also 2-fold. Because it is possible to have multiple {001} domains in experimental samples for 1q and 2q, and if the domain size is much smaller than the ARPES laser beam spot, the overlapped 2-fold SSP and bulk FS from the mixed {001} domains become effectively 4-fold SSP with a bulk FS background (see Fig.3(o) and (p) for a direct average), which is in a good agreement with ARPES data showing that the hole Fermi arc-like feature and surface electron pocket have an underneath bulk FS background. Thus, this observation can be used to exclude 3q, because 3q only has one (001) domain (Fig.3(k)) without the bulk FS background underneath the 4-fold SSP. The existence of 1q or 2q different {001} domains with the 2-fold SSP has been observed in the sibling compound NdSb[45].

The detailed comparison of SSP dispersion to ARPES data will next be made with 2q (001) in Fig.4. As highlighted by the orange box in Fig.4(a), the spin texture of the SSP hole arc and electron pockets are plotted in Fig.4(b). The orange arrows show the in-plane components of the expectation value of electron spin weighted by the SFS at the specific $k$-point and energy (see Eqn.(32) in Ref.[46]). The opposite spin texture between the SSP hole arc and electron pocket is persistent when moving to more negative energies with the SS electron pocket shrinking in size, which is in very good agreement with the spin-polarized ARPES data[22].

To compare with ARPES data[22] side-by-side and analyze the SSP away from the high symmetry directions, the same two sets of six cuts each as for ARPES (see Fig.3 of Ref.[22]) are made with black and white dashed lines in Fig.4(a) to calculate the surface spectral function on 2q (001) in detail. With the black vertical dashed cuts to focus on half of the hole Fermi arc-like feature, the spectral function in Fig.4(e) shows the hole arc (the lower SS) moves to more negative energy and also gradually merges with the bulk hole band when shifted away from $k_x$=0 along the $\bar{\Gamma}$-$\bar{X}$ direction in Cut No.1-3. These behaviors of the hole Fermi arc-like feature are in good agreement with those observed in Fig.4(c). When shifted further, both calculation and ARPES also agree on the arc tip still appearing at Cut No.4, while disappearing at Cut No.5-6 with only 3D bulk bands left behind. For the cuts with the white dashed lines rotated 45 degree and moved away from the $\bar{\Gamma}$-$\bar{M}$ direction in Fig.4(a) to survey both the hole Fermi arc-like feature and SS electron pocket, the emerging arc tip at Cut No.2 in Fig.4(f), the gradual emergence and shift of the asymmetric splitting between the upper and lower SS at Cut No.3-6 also agree very well with the ARPES data in Fig.4(d). The enveloping of the lower SS to bulk hole band and that of upper SS to bulk electron band are also evident. The similarly good agreement to ARPES data for the 2-fold SSP on 2q (010) and 1q (110) with the same sets of cuts can be found in Supplementary Note 4 and Fig.S3. These good agreement between the calculated unconventional SSP and the ARPES data indicates a general rule that the unconventional SSP emerges in NdBi along the {001} directions having the AFM band-folding bulk hybridization gap for the different type-I AFM multi-q structures. This common theme is the main result of the paper.

## II-C. Nodal features in multi-q structures

As mentioned above, in addition to the unconventional SSP, type-I AFM multi-q NdBi also have nodal features. We will show that the unconventional SSP does not arise from the WPs in 3q or the DPs in 2q near the $E_F$. For 3q with the all-in-all-out magnetic configuration (first predicted Weyl semimetal in iridates[47]), we found 16 WPs in two sets along the $\Gamma$-$R$ direction (see Fig.5(a)) with the momentum-energy of (±0.0875, ±0.0875, ±0.0875 1/Å, $E_F$+12.6 meV) and (±0.1133, ±0.1133, ±0.1133 1/Å, $E_F$+14.9 meV), respectively. In contrast to the axion insulator in NpBi with the same 3q structure as analyzed in the Xu et al.[12], the even number of WPs in half of the BZ is also compatible with $Z_4$=2 for NdBi 3q here. The Berry curvatures near a pair of these WPs are plotted in the vertical and horizontal planes in Fig.5(b) and (c) to confirm the opposite chirality of ±1. When projected on (001) surface, the WPs of opposite chirality overlap and usually Fermi arcs are not expected. But as shown by the SFS around the WP binding energy of $E_F$+0.014 eV in Fig.5(d) and (e), there are small protrusions along the $\bar{\Gamma}$-$\bar{M}$ direction connecting the hole SS on both the top and bottom surfaces. After zooming in Fig.5(f) and (g) with the surface-only contribution, it turns out that the hole SS connects the projections of outer WPs (green circles), but not the inner WPs (black squares). The usual explanation for the disappearance of open Fermi arcs when two WPs of the opposite chirality overlap has the assumption that the open Fermi arcs connect in the direction of the shortest path between the WP projections. When the WP projections of opposite chirality move closer and overlap, such as the black squares, Fermi arcs disappear. However, as shown in the earlier study[48], the open Fermi arc connectivity of the WP projections are not restricted to only one direction and can actually change to the opposite direction depending on the surrounding SS. In such a scenario, the overlap of WP projections of opposite chirality can give a full circle for the Fermi arcs, which is seen here for the case of the outer set of WPs in NdBi 3q. With projection to the 3q (111) surface as shown in Fig.S4 of Supplementary Note 5, there are Fermi arcs connecting the projections of the WPs without overlap of the opposite chirality.

For the DPs in 2q, the bulk band structure is plotted in Fig.S5(a) of Supplementary Note 6 along the $\Gamma$-$Z$ direction, where the top valence and bottom conduction bands cross twice with different orbital content. The two pairs of DPs have the momentum-energy of (0, 0, ±0.1501 1/Å, $E_F$+73.0 meV) and (0, 0, ±0.4330 1/Å, $E_F$–177.5 meV), respectively. The upper DP has a switch of orbital content between Bi $p_z$ and $p_x$, while the lower DP has the switch between Nd $d_{xy}$ and Bi $p_x$ orbitals. These DPs are protected by the 4-fold screw axis. To search for any topological SS signature of the DPs, the side (010) surface spectral function along the $\bar{\Gamma}$-$\bar{Z}$ direction is presented in Fig.S5(b). But the SS around the binding energy of the upper DP at $E_F$+0.073 eV does not originate from the DP projection. For the lower DP at $E_F$–0.178 eV, two SS follow the narrow band gap from $E_F$ to $E_F$–0.1 eV moving toward the DP, but they merge into the bulk band projection before reaching the DP. The lack of the closed Fermi arcs connecting two DP projections in 2q is also seen in the SFS in Fig.S5(c) and (d) at the binding energies of these DPs. Although the closed Fermi arcs connecting two DP projections are often found in some other Dirac semimetals, they are not guaranteed with topological protection and are fragile as discussed before in a theoretical model[49] and also found for PtBi$_2$[50]. Our results for NdBi 2q provide another example regarding the fragility of the Fermi arcs for DPs.

From the perspective of using symmetry-based indicators and topological invariants to diagnose topological states for magnetic space groups, a previous theory work[12] has noticed that the parity-based index $Z_4$=2 can correspond to different topological states including both gapped and gapless phases. Although NpBi with 3q has been predicted[12] as an axion insulator with gapped surface Dirac cones and chiral hinge states, our current study on multi-q NdBi realizes several other compatible scenarios with $Z_4$=2, including a Weyl semimetal with an even number of WPs in half of the bulk BZ for 3q and a Dirac semimetal for 1q and 2q. The unconventional SSP spanning the AFM band-folding hybridization gap along the {001} directions remains very robust for all these changes in multi-q structures to give the key features of unconventional surface hole Fermi arc-like features and elliptical electron pocket as observed in ARPES. Thus, besides explaining of the unconventional SSP in NdBi below $T_N$, our study also expands the variety of topological features that are compatible with the parity-based index $Z_4$=2. However for the SSP, finding the exact topological invariants to associate with is still an open question.

We have also changed the U and J values in the DFT calculations. The DFT total energy differences with the fully relaxed unit cells in PBEsol+U+SOC are plotted in Fig.S6(a) of Supplementary Note 7 as a function of U. The 2q becomes the most stable phase when the U value is larger than 9.6 eV. We have calculated the 2q surface band structures with a U value of 9.8 eV on (001) and (010) in Fig.S6, and found the existence and connectivity of the unconventional SSP in the AFM band-folding hybridization gap remain the same. But the splitting between the SSP is reduced as the Nd 4*f* bands are moved further away from the $E_F$ with higher U value. The simultaneous description of the right size of SSP splitting and total energy difference is challenging for the DFT+U method. The temperature dependence of the SSP splitting can be more complicated and needs further theory efforts.

## III. Conclusion

We find that the type-I anti-ferromagnetic multi-q structures, 1q, 2q and 3q of NdBi, can host unconventional surface state pairs (SSP) that are consistent with the ARPES data around the $\bar{\varGamma}$ point below $T_N$. These unconventional SS appear in pairs and reside in the bulk hybridization gap region formed by the AFM band-folding of the band inversion region along the {001} directions. The lower SS forms the Fermi arc-like features connected with the bulk hole pocket and the upper SS forms an elliptical electron pocket with a cusp toward the $\bar{X}$ point, agreeing very well with the experiment. We also find NdBi 2q is a Dirac semimetal with two pairs of Dirac points (DPs) along the $\varGamma$-$Z$ direction, but without the closed Fermi arcs connecting the DP projections. In contrast, NdBi 3q is a Weyl semimetal with two sets of 16 Weyl points (WPs) along the $\varGamma$-$R$ direction. The projections of the outer set are connected by the hole SS arcs in a full circle, despite that the WP projections are overlapped with opposite chirality. Our results also expand a variety of topological nodal features that are compatible with the parity-based $Z_4$=2.

## IV. Methods

All density functional theory[25, 26] (DFT) calculations with spin-orbit coupling (SOC) were performed with the PBE[51] exchange-correlation functional using a plane-wave basis set and projector augmented wave method[52], as implemented in the Vienna Ab-initio Simulation Package (VASP)[53, 54]. Using maximally localized Wannier functions[55, 56], tight-binding models were constructed to reproduce closely the band structure including SOC within $E_F$±1eV with Nd *s-d-f* and Bi *p* orbitals. The surface spectral function and 2D Fermi surface (FS) were calculated with the surface Green's function methods[57, 58] as implemented in WannierTools[46]. In the DFT calculations, we used a kinetic energy cutoff of 253 eV, $\Gamma$-centered Monkhorst-Pack[59] (8×8×8) *k*-point mesh, and a Gaussian smearing of 0.05 eV. For band structure calculations of NdBi, we have used the experimental structural parameter[60] of 6.41 Å. To account for the strongly localized Nd 4*f* orbitals, an onsite Hubbard-like[61] U=6.3 eV and J=0.7 eV have been used. Our DFT+U+SOC calculation gives a spin moment of 2.7 $\mu_B$ and an orbital moment of 5.8 $\mu_B$ in the opposite direction, resulting in a total magnetic moment of 3.1 $\mu_B$ on Nd, agreeing very well with the experimental data[27] of 3.1±0.2 $\mu_B$.

## Acknowledgements

This work was supported in the Center for the Advancement of Topological Semimetals, an Energy Frontier Research Center funded by the U.S. Department of Energy Office of Science, Office of Basic Energy Sciences through the Ames Laboratory under its Contract No. DE-AC02-07CH11358. Part of this work was also supported by the U.S. Department of Energy, Office of Basic Energy Science, Division of Materials Sciences and Engineering. The research was performed at the Ames Laboratory. Ames Laboratory is operated for the U.S. Department of Energy by Iowa State University under Contract No. DE-AC02-07CH11358. R.-J. S. in addition acknowledges funding via the Marie Sklodowska-Curie programme [EC Grant No. 842901] and the Winton programme as well as Trinity College at the University of Cambridge.

**Author Contributions**: P.C.C. and A.K. initiated and conceived the work on RPn. L.-L.W. designed and performed the ab initio calculations with inputs from J.A., R.-J.S., B.G.U and R.J.M. A.K., P.C.C., B.S., Y.K., A.S. and B.K. provided experimental inputs. All authors discussed the results and contributed to the final manuscript.

**Competing Interests**: The authors declare no competing interests.

**Data Availability**: The data that support the findings of this study are available from the corresponding author upon reasonable request.

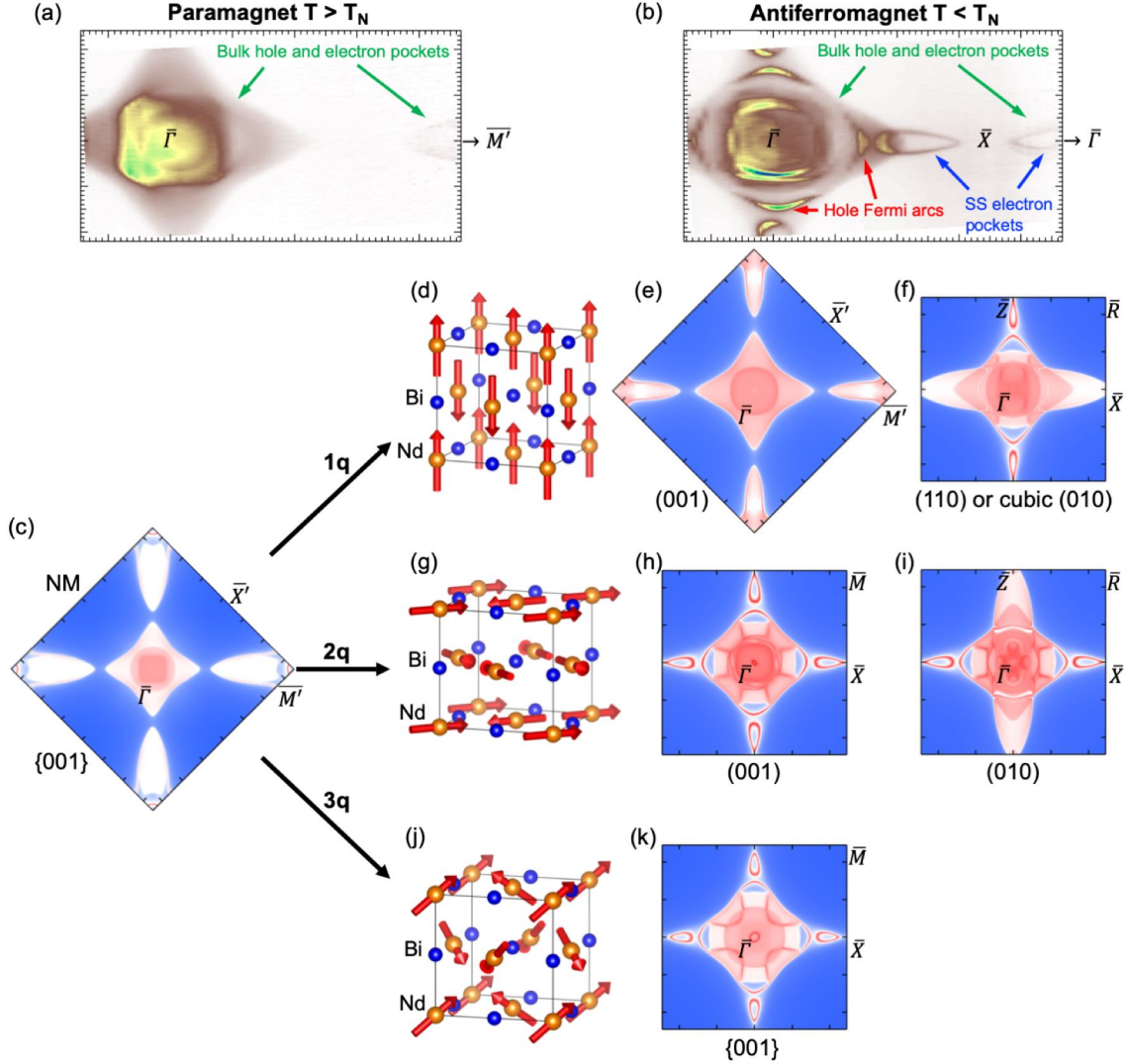

Figure 1. 2D surface Fermi surface (SFS) of NdBi. Angle-resolved photo-electron spectroscopy (ARPES) at (a) T>$T_N$ and (b) T< $T_N$ with the unconventional surface state (SS) of hole Fermi arcs and SS electron pockets. The calculated SFS on (001) for (c) non-magnetic (NM) and anti-ferromagnetic (AFM) with (d)-(f) 1q, (g)-(i) 2q and (j)-(k) 3q wave vectors on different surfaces as labeled. The multi-q magnetic structures are shown in (d), (g) and (j). The orange (blue) spheres stand for Nd (Bi) atoms and the red arrows show the directions of Nd magnetic moments. The surface Brillouin zone for NM and 1q (001) is twice as large and rotated by 45 degree with respect to that for 2q and 3q (see Fig.2(b) for labels of high symmetry points). The 1q tetragonal (110) vertical diagonal surface in (f) is equivalent to the cubic (010) side surface. Blue, white and red colors stand for low, medium, and high density of states.

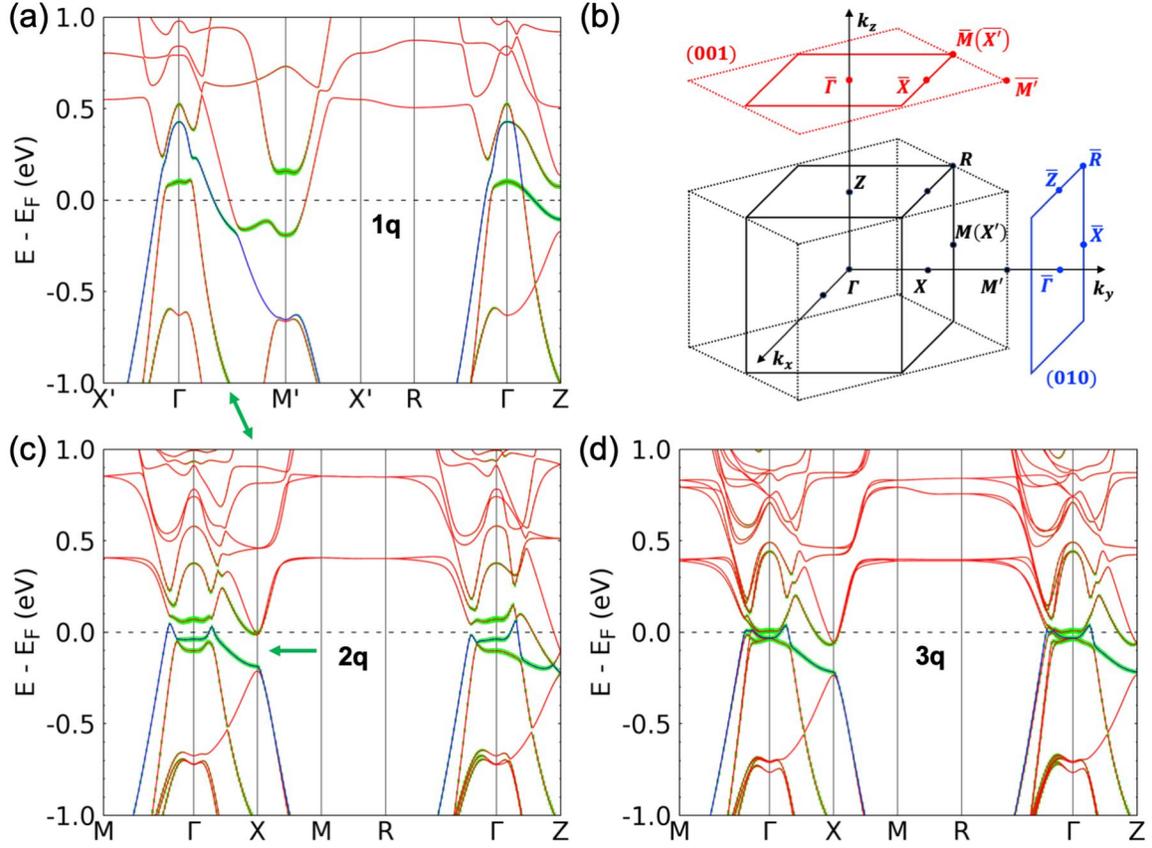

Figure 2. Calculated bulk band structure of NdBi. (a) 1q, (c) 2q and (d) 3q are plotted with respect to the Fermi energy ($E_F$) at 0.0 eV. The top valence band is in blue and the green shade stands for the projection of Bi $p$ orbitals. (b) 3D bulk Brillouin zone (BZ) of 1q (dashed lines for the tetragonal unit cell) and that of 2q and 3q (solid lines for the cubic unit cell), and the 2D surface Brillouin zone (SBZ) of (001) and (010) with the high-symmetry points labeled. The green vertical arrow in (a) points to band-folding at the $X$ point, half-way along the $\Gamma$-$M'$ direction for 1q. The green horizontal arrow in (c) points to hybridization gap opening upon band-folding along the $\Gamma$-$X$ direction for 2q.

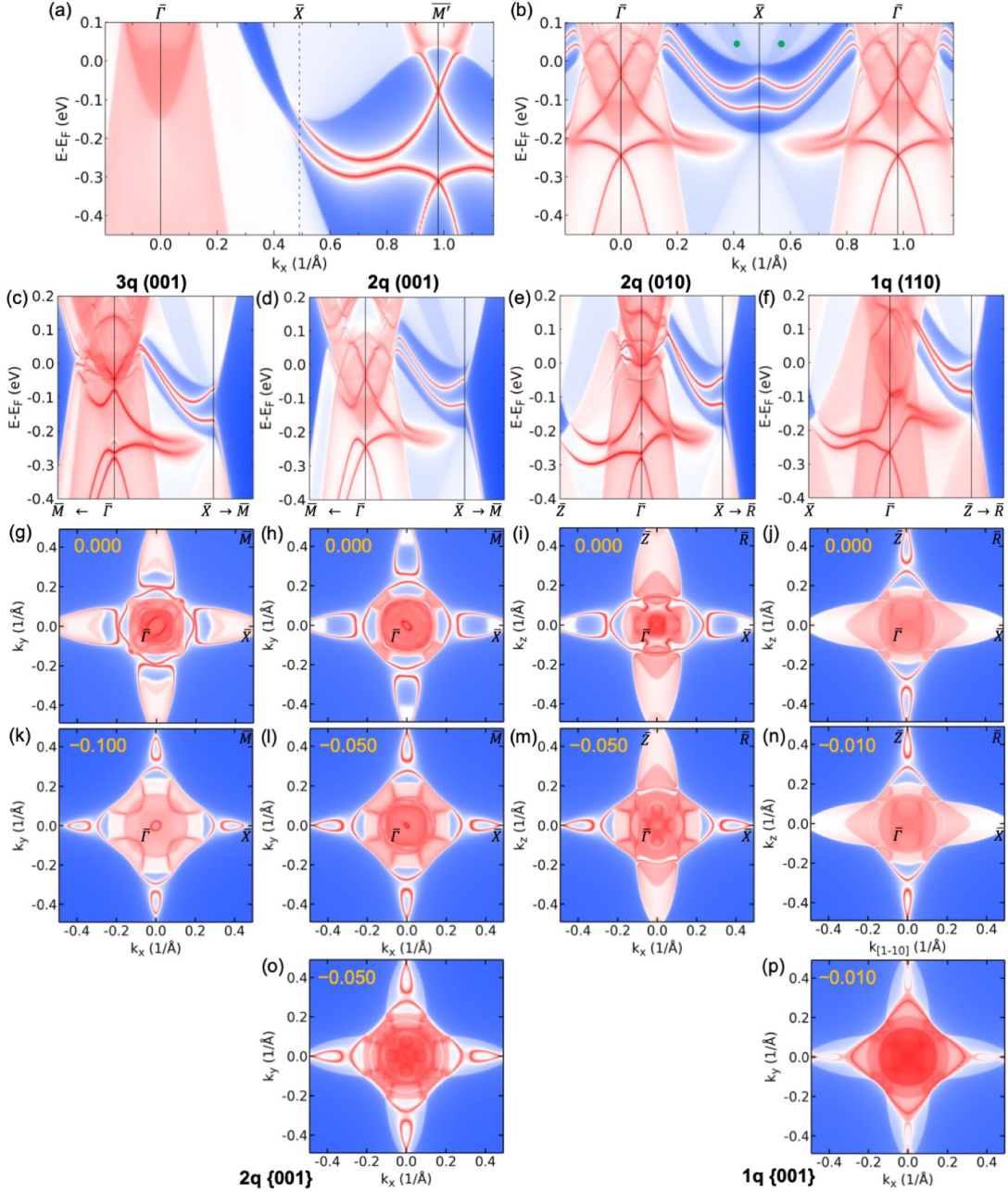

Figure 3. Surface spectral functions of NdBi. (a) nonmagnetic (NM) (001) along $\bar{\Gamma}$-$\bar{X}$-$\overline{M'}$ and (b) 2q (001) along the equivalent direction of $\bar{\Gamma}$-$\bar{X}$-$\bar{\Gamma}$. The green dots indicate the original band inversion region along $\bar{\Gamma}$-$\bar{X}$, just above the bulk hybridization gap and the associated unconventional surface state pairs. Surface spectral functions on (c) 3q (001), (d) 2q (001), (e) 2q (010) and (f) 1q (110) along other directions. 2D surface Fermi surfaces at two different energies as labeled in eV with respect to Fermi energy ($E_F$) for four surfaces: (g) and (k) for 3q (001), (h) and (l) for 2q (001), (i) and (m) for 2q (010), and (j) and (n) for 1q (110), respectively. (o) and (p) are the averaged cubic {001} domains for 2q and 1q, respectively. Blue, white and red colors stand for low, medium, and high density of states.

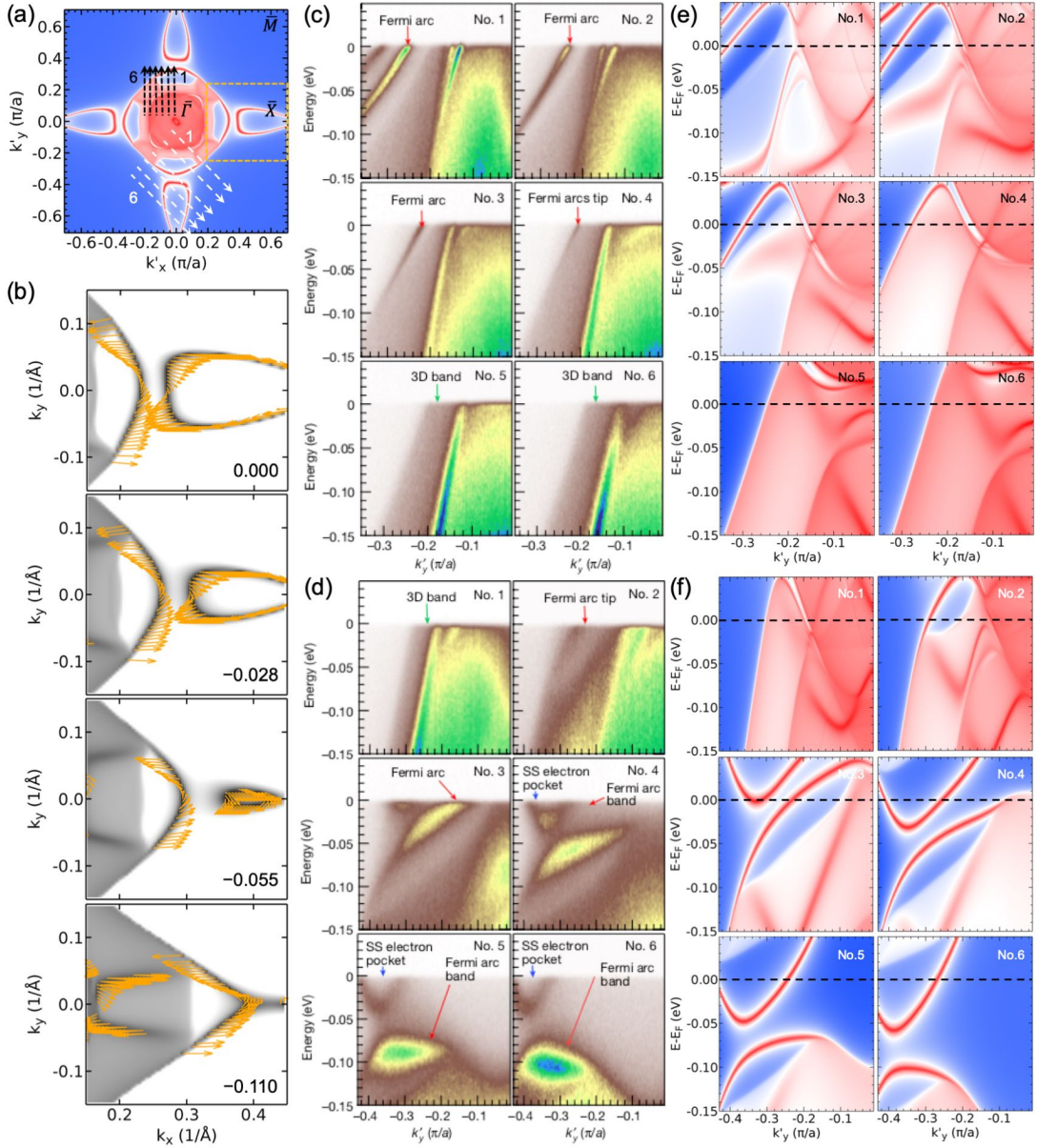

Figure 4. Comparison between NdBi 2q (001) surface states and angle-resolved photo-electron spectroscopy (ARPES). (a) 2D surface Fermi surface (SFS) on 2q (001) labeled with two sets of cuts in black and white dashed lines, and an orange box focusing on the surface state pairs. (b) SFS zoomed in the orange box of (a) for the surface hole Fermi arc and electron pocket with the spin texture (in-plane components of the expectation value of spin weighted by the SFS) indicated by the orange arrows. The different energies are labeled in eV. (c) and (e) Comparison between ARPES and the calculated spectral functions at the cuts (black dashed lines in (a)) of 0.03, 0.08, 0.12, 0.14, 0.18 and 0.20 in the unit of the nonmagnetic surface Brillouin zone. (d) and (f) Comparison at cuts in the other direction (white dashed lines in (a)) at 0.03, 0.09, 0.17, 0.22, 0.24 and 0.27. All energies are downward shifted by 0.010 eV from the Fermi energy ($E_F$).

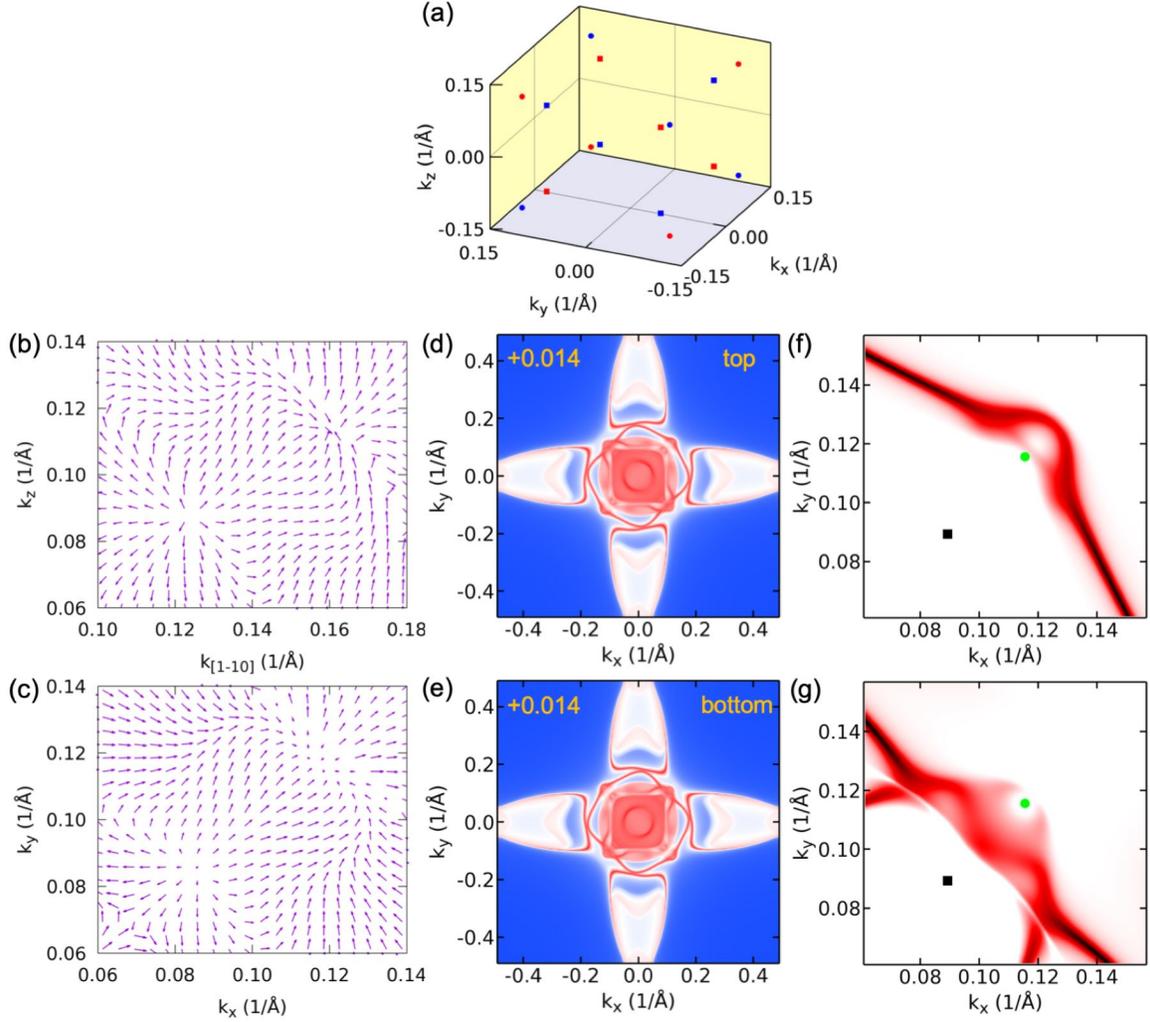

Figure 5. Weyl points (WPs) in NdBi 3q. (a) Locations of the two sets of 16 WPs with the momentum-energy of (±0.0875, ±0.0875, ±0.0875 1/Å, $E_F$+12.6 meV) and (±0.1133, ±0.1133, ±0.1133 1/Å, $E_F$+14.9 meV), respectively, where $E_F$ is the Fermi energy. Berry curvature near the WP pair on the (b) vertical $k_{xy}$-$k_z$ plane and (c) horizontal $k_x$-$k_y$ plane at $k_z$=0.10. Surface Fermi surface near the WP energy at $E_F$+0.014 eV on the (d) top and (e) bottom (001) surfaces. The corresponding surface-only contributions zoomed on the (f) top and (g) bottom (001) surface with the locations of the WP projections labeled by the black square and green circle.

# References

1. C.-G. Duan, R. F. Sabirianov, W. N. Mei, P. A. Dowben, S. S. Jaswal, E. Y. Tsymbal, Electronic, magnetic and transport properties of rare-earth monopnictides. *Journal of Physics: Condensed Matter* **19**, 315220 (2007).
2. M. Zeng, C. Fang, G. Chang, Y.-A. Chen, T. Hsieh, A. Bansil, H. Lin, L. Fu, Topological semimetals and topological insulators in rare earth monopnictides. arXiv:1504.03492 (2015).
3. Y. Wu, T. Kong, L. L. Wang, D. D. Johnson, D. X. Mou, L. N. Huang, B. Schrunk, S. L. Bud'ko, P. C. Canfield, A. Kaminski, Asymmetric mass acquisition in LaBi: Topological semimetal candidate. *Phys Rev B* **94**, 081108(R) (2016).
4. Y. Wu, Y. Lee, T. Kong, D. X. Mou, R. Jiang, L. A. Huang, S. L. Bud'ko, P. C. Canfield, A. Kaminski, Electronic structure of RSb (R = Y, Ce, Gd, Dy, Ho, Tm, Lu) studied by angle-resolved photoemission spectroscopy. *Phys Rev B* **96**, 035134 (2017).
5. N. Alidoust, A. Alexandradinata, S.-Y. Xu, I. Belopolski, S. K. Kushwaha, M. Zeng, M. Neupane, G. Bian, C. Liu, D. S. Sanchez, P. P. Shibayev, H. Zheng, L. Fu, A. Bansil, H. Lin, R. J. Cava, M. Zahid Hasan, A new form of (unexpected) Dirac fermions in the strongly-correlated cerium monopnictides. arXiv:1604.08571 (2016).
6. J. Nayak, S. C. Wu, N. Kumar, C. Shekhar, S. Singh, J. Fink, E. E. D. Rienks, G. H. Fecher, S. S. P. Parkin, B. H. Yan, C. Felser, Multiple Dirac cones at the surface of the topological metal LaBi. *Nat Commun* **8**, 13942 (2017).
7. P. Li, Z. Wu, F. Wu, C. Cao, C. Guo, Y. Wu, Y. Liu, Z. Sun, C.-M. Cheng, D.-S. Lin, F. Steglich, H. Yuan, T.-C. Chiang, Y. Liu, Tunable electronic structure and surface states in rare-earth monobismuthides with partially filled $f$ shell. *Phys Rev B* **98**, 085103 (2018).
8. K. Kuroda, M. Ochi, H. S. Suzuki, M. Hirayama, M. Nakayama, R. Noguchi, C. Bareille, S. Akebi, S. Kunisada, T. Muro, M. D. Watson, H. Kitazawa, Y. Haga, T. K. Kim, M. Hoesch, S. Shin, R. Arita, T. Kondo, Experimental Determination of the Topological Phase Diagram in Cerium Monopnictides. *Phys Rev Lett* **120**, 086402 (2018).
9. X. Duan, F. Wu, J. Chen, P. Zhang, Y. Liu, H. Yuan, C. Cao, Tunable electronic structure and topological properties of LnPn (Ln=Ce, Pr, Sm, Gd, Yb; Pn=Sb, Bi). *Communications Physics* **1**, 71 (2018).
10. D.-C. Ryu, J. Kim, K. Kim, C.-J. Kang, J. D. Denlinger, B. I. Min, Distinct topological properties in Ce monopnictides having correlated $f$ electrons: CeN vs. CeBi. *Physical Review Research* **2**, 012069 (2020).
11. H. Watanabe, H. C. Po, A. Vishwanath, Structure and topology of band structures in the 1651 magnetic space groups. *Science Advances* **4**, eaat8685 (2018).

12. Y. Xu, L. Elcoro, Z.-D. Song, B. J. Wieder, M. G. Vergniory, N. Regnault, Y. Chen, C. Felser, B. A. Bernevig, High-throughput calculations of magnetic topological materials. *Nature* **586**, 702-707 (2020).

13. A. Bouhon, G. F. Lange, R.-J. Slager, Topological correspondence between magnetic space group representations and subdimensions. *Phys Rev B* **103**, 245127 (2021).

14. S. X. M. Riberolles, T. V. Trevisan, B. Kuthanazhi, T. W. Heitmann, F. Ye, D. C. Johnston, S. L. Bud'ko, D. H. Ryan, P. C. Canfield, A. Kreyssig, A. Vishwanath, R. J. McQueeney, L. L. Wang, P. P. Orth, B. G. Ueland, Magnetic crystalline-symmetry-protected axion electrodynamics and field-tunable unpinned Dirac cones in $EuIn_2As_2$. *Nat Commun* **12**, 999 (2021).

15. L.-L. Wang, H. C. Po, R.-J. Slager, A. Vishwanath, Topological descendants of a multicritical Dirac semimetal with magnetism and strain. *Phys Rev B* **104**, 165107 (2021).

16. J. Kruthoff, J. de Boer, J. van Wezel, C. L. Kane, R.-J. Slager, Topological Classification of Crystalline Insulators through Band Structure Combinatorics. *Phys Rev X* **7**, 041069 (2017).

17. C. Y. Guo, C. Cao, M. Smidman, F. Wu, Y. J. Zhang, F. Steglich, F. C. Zhang, H. Q. Yuan, Possible Weyl fermions in the magnetic Kondo system CeSb. *Npj Quantum Mater* **2**, 39 (2017).

18. Z. Li, D.-D. Xu, S.-Y. Ning, H. Su, T. Iitaka, T. Tohyama, J.-X. Zhang, Predicted Weyl fermions in magnetic GdBi and GdSb. *International Journal of Modern Physics B* **31**, 1750217 (2017).

19. Z. Huang, C. Lane, C. Cao, G.-X. Zhi, Y. Liu, C. E. Matt, B. Kuthanazhi, P. C. Canfield, D. Yarotski, A. J. Taylor, J.-X. Zhu, Prediction of spin polarized Fermi arcs in quasiparticle interference in CeBi. *Phys Rev B* **102**, 235167 (2020).

20. H. Oinuma, S. Souma, K. Nakayama, K. Horiba, H. Kumigashira, M. Yoshida, A. Ochiai, T. Takahashi, T. Sato, Unusual change in the Dirac-cone energy band upon a two-step magnetic transition in CeBi. *Phys Rev B* **100**, 125122 (2019).

21. P. Li, Z. Wu, F. Wu, C. Guo, Y. Liu, H. Liu, Z. Sun, M. Shi, F. Rodolakis, J. L. McChesney, C. Cao, H. Yuan, F. Steglich, Y. Liu, Large Fermi surface expansion through anisotropic mixing of conduction and f electrons in the semimetallic Kondo lattice CeBi. *Phys Rev B* **100**, 155110 (2019).

22. B. Schrunk, Y. Kushnirenko, B. Kuthanazhi, J. Ahn, L.-L. Wang, E. O'Leary, K. Lee, A. Eaton, A. Fedorov, R. Lou, V. Voroshnin, O. J. Clark, J. Sánchez-Barriga, S. L. Bud'ko, R.-J. Slager, P. C. Canfield, A. Kaminski, Emergence of Fermi arcs due to magnetic splitting in an antiferromagnet. *Nature* **603**, 610-615 (2022).

23. M. R. Norman, H. Ding, M. Randeria, J. C. Campuzano, T. Yokoya, T. Takeuchi, T. Takahashi, T. Mochiku, K. Kadowaki, P. Guptasarma, D. G. Hinks, Destruction of the Fermi surface in underdoped high-Tc superconductors. *Nature* **392**, 157-160 (1998).

24. S. Y. Xu, I. Belopolski, N. Alidoust, M. Neupane, G. Bian, C. L. Zhang, R. Sankar, G. Q. Chang, Z. J. Yuan, C. C. Lee, S. M. Huang, H. Zheng, J. Ma, D. S. Sanchez, B. K. Wang, A. Bansil, F. C. Chou, P. P. Shibayev, H. Lin, S. Jia, M. Z. Hasan, Discovery of a Weyl fermion semimetal and topological Fermi arcs. *Science* **349**, 613-617 (2015).

25. P. Hohenberg, W. Kohn, Inhomogeneous Electron Gas. *Phys. Rev.* **136**, B864-B871 (1964).

26. W. Kohn, L. J. Sham, Self-Consistent Equations Including Exchange and Correlation Effects. *Phys. Rev.* **140**, A1133-A1138 (1965).

27. N. Nereson, G. Arnold, Magnetic Properties of CeBi, NdBi, TbBi, and DyBi. *Journal of Applied Physics* **42**, 1625-1627 (1971).

28. J. Rossat-Mignod, P. Burlet, S. Quezel, O. Vogt, Magnetic ordering in cerium and uranium monopnictides. *Physica B+C* **102**, 237-248 (1980).

29. P. Burlet, F. Bourdarot, J. Rossat-Mignod, J. P. Sanchez, J. C. Spirlet, J. Rebizant, O. Vogt, Neutron diffraction study of the magnetic ordering in NpBi. *Physica B: Condensed Matter* **180-181**, 131-132 (1992).

30. Y. Yamamoto, T. Nagamiya, Spin Arrangements in Magnetic Compounds of the Rocksalt Crystal Structure. *J Phys Soc Jpn* **32**, 1248-1261 (1972).

31. J. P. Ader, Magnetic order in the frustrated Heisenberg model for the fcc type-I configuration. *Phys Rev B* **65**, 014411 (2001).

32. S.-S. Diop, G. Jackeli, L. Savary, Anisotropic exchange and noncollinear antiferromagnets on a noncentrosymmetric fcc half-Heusler structure. *Phys Rev B* **105**, 144431 (2022).

33. T. H. R. Skyrme, A unified field theory of mesons and baryons. *Nuclear Physics* **31**, 556-569 (1962).

34. N. Nagaosa, Y. Tokura, Topological properties and dynamics of magnetic skyrmions. *Nature Nanotechnology* **8**, 899-911 (2013).

35. A. N. Bogdanov, U. K. Rößler, Chiral Symmetry Breaking in Magnetic Thin Films and Multilayers. *Phys Rev Lett* **87**, 037203 (2001).

36. B. Binz, A. Vishwanath, V. Aji, Theory of the Helical Spin Crystal: A Candidate for the Partially Ordered State of MnSi. *Phys Rev Lett* **96**, 207202 (2006).

37. T. Okubo, S. Chung, H. Kawamura, Multiple-q States and the Skyrmion Lattice of the Triangular-Lattice Heisenberg Antiferromagnet under Magnetic Fields. *Phys Rev Lett* **108**, 017206 (2012).

38. P. Milde, D. Köhler, J. Seidel, L. M. Eng, A. Bauer, A. Chacon, J. Kindervater, S. Mühlbauer, C. Pfleiderer, S. Buhrandt, C. Schütte, A. Rosch, Unwinding of a Skyrmion Lattice by Magnetic Monopoles. *Science* **340**, 1076-1080 (2013).

39. S. Seo, S. Hayami, Y. Su, S. M. Thomas, F. Ronning, E. D. Bauer, J. D. Thompson, S.-Z. Lin, P. F. S. Rosa, Spin-texture-driven electrical transport in multi-Q antiferromagnets. *Communications Physics* **4**, 58 (2021).

40. X. R. Wang, X. C. Hu, H. T. Wu, Stripe skyrmions and skyrmion crystals. *Communications Physics* **4**, 142 (2021).

41. X.-C. Hu, H.-T. Wu, X. R. Wang, A theory of skyrmion crystal formation. *Nanoscale* **14**, 7516-7529 (2022).

42. S. Hayami, Multifarious skyrmion phases on a trilayer triangular lattice. *Phys Rev B* **105**, 184426 (2022).

43. K. Shimizu, S. Okumura, Y. Kato, Y. Motome, Phase degree of freedom and topology in multiple-Q spin textures. *Phys Rev B* **105**, 224405 (2022).

44. S. Okumura, S. Hayami, Y. Kato, Y. Motome, Magnetic Hedgehog Lattice in a Centrosymmetric Cubic Metal. *J Phys Soc Jpn* **91**, 093702 (2022).

45. Y. Kushnirenko, B. Schrunk, B. Kuthanazhi, L.-L. Wang, J. Ahn, E. O'Leary, A. Eaton, S. L. Bud'ko, R.-J. Slager, P. C. Canfield, A. Kaminski, Rare-earth monopnictides: Family of antiferromagnets hosting magnetic Fermi arcs. *Phys Rev B* **106**, 115112 (2022).

46. Q. Wu, S. Zhang, H.-F. Song, M. Troyer, A. A. Soluyanov, WannierTools: An open-source software package for novel topological materials. *Computer Physics Communications* **224**, 405-416 (2018).

47. X. G. Wan, A. M. Turner, A. Vishwanath, S. Y. Savrasov, Topological semimetal and Fermi-arc surface states in the electronic structure of pyrochlore iridates. *Phys Rev B* **83**, 205101 (2011).

48. L.-L. Wang, Expansive open Fermi arcs and connectivity changes induced by infrared phonons in $ZrTe_5$. *Phys Rev B* **103**, 075105 (2021).

49. M. Kargarian, M. Randeria, Y.-M. Lu, Are the surface Fermi arcs in Dirac semimetals topologically protected? *Proceedings of the National Academy of Sciences* **113**, 8648 (2016).

50. Y. Wu, N. H. Jo, L.-L. Wang, C. A. Schmidt, K. M. Neilson, B. Schrunk, P. Swatek, A. Eaton, S. L. Bud'ko, P. C. Canfield, A. Kaminski, Fragility of Fermi arcs in Dirac semimetals. *Phys Rev B* **99**, 161113 (2019).

51. J. P. Perdew, K. Burke, M. Ernzerhof, Generalized gradient approximation made simple. *Phys Rev Lett* **77**, 3865-3868 (1996).

52. P. E. Blöchl, Projector Augmented-Wave Method. *Phys Rev B* **50**, 17953-17979 (1994).

53. G. Kresse, J. Furthmuller, Efficient Iterative Schemes for Ab initio Total-Energy Calculations Using a Plane-Wave Basis Set. *Phys Rev B* **54**, 11169-11186 (1996).

54. G. Kresse, J. Furthmuller, Efficiency of Ab-initio Total Energy Calculations for Metals and Semiconductors Using a Plane-Wave Basis Set. *Comp Mater Sci* **6**, 15-50 (1996).

55. N. Marzari, D. Vanderbilt, Maximally localized generalized Wannier functions for composite energy bands. *Phys Rev B* **56**, 12847-12865 (1997).

56. I. Souza, N. Marzari, D. Vanderbilt, Maximally localized Wannier functions for entangled energy bands. *Phys Rev B* **65**, 035109 (2001).

57. M. P. L. Sancho, J. M. L. Sancho, J. Rubio, Quick Iterative Scheme for the Calculation of Transfer-Matrices - Application to Mo(100). *J Phys F Met Phys* **14**, 1205-1215 (1984).

58. M. P. L. Sancho, J. M. L. Sancho, J. Rubio, Highly Convergent Schemes for the Calculation of Bulk and Surface Green-Functions. *J Phys F Met Phys* **15**, 851-858 (1985).

59. H. J. Monkhorst, J. D. Pack, Special Points for Brillouin-Zone Integrations. *Phys Rev B* **13**, 5188-5192 (1976).

60. M. N. Abdusalyamova, O. I. Rakhmatov, K. S. Shokirov, Properties of monobismuthides of rare-earth metals of the cerium subgroup. *Russ. Metall.* **1**, 183-185 (1988).

61. A. I. Liechtenstein, V. I. Anisimov, J. Zaanen, Density-functional theory and strong interactions: Orbital ordering in Mott-Hubbard insulators. *Phys Rev B* **52**, R5467-R5470 (1995).

# Supplementary Information

# Unconventional Surface State Pairs in a High-Symmetry Lattice with Anti-ferromagnetic Band-folding

Lin-Lin Wang[1*], Junyeong Ahn[2], Robert-Jan Slager[3], Yevhen Kushnirenko[1], Benjamin G. Ueland[1], Aashish Sapkota[1], Benjamin Schrunk[1], Brinda Kuthanazhi[4], Robert J. McQueeney[1,4], Paul C. Canfield[1,4] and Adam Kaminski[1,4]

[1]Ames National Laboratory, U.S. Department of Energy, Ames, IA 50011, USA
[2]Department of Physics, Harvard University, Cambridge, MA 02138, USA
[3]TCM Group, Cavendish Laboratory, University of Cambridge, J. J. Thomson Avenue, Cambridge CB3 0HE, UK
[4]Department of Physics, Iowa State University, Ames, IA 50011, USA

*llw@amelsab.gov

**Supplementary Note 1: NM (001) bulk and surface band structures**

**Supplementary Note 2: Multi-q magnetic space groups and bulk band structures**

**Supplementary Note 3: 1q surface band structures on other surfaces**

**Supplementary Note 4: Comparison for 2-fold SSP of 2q and 1q to ARPES**

**Supplementary Note 5: 3q (111) surface projection**

**Supplementary Note 6: Dirac points in 2q**

**Supplementary Note 7: DFT total energy with relaxation**

## Supplementary Note 1: NM (001) bulk and surface band structures

Bulk band structure of non-magnetic (NM) NdBi is plotted in Fig.S1(a). The band inversion with anti-crossing along the $\Gamma$-$X$ direction is between the Bi $p$ and Nd $d$ orbitals for hosting a strong topological insulator (TI) state with the Fu-Kane index of (1;000). As shown in Fig.S1(b), the topological surface states emerge from the band inversion region and disperse toward the $\overline{M'}$ point and form two gapped surface Dirac points at $\overline{M'}$. Upon the simple band-folding without hybridization by using the larger unit cell, the same as 3q, the topological surface states associated with the strong TI are folded to the $\overline{\Gamma}$ point and overlapped with the bulk band projections in the folded tetragonal SBZ (as zoomed in Fig.S1(c)). But there is no hybridization gap opening or unconventional surface state pairs as seen in Fig.3.

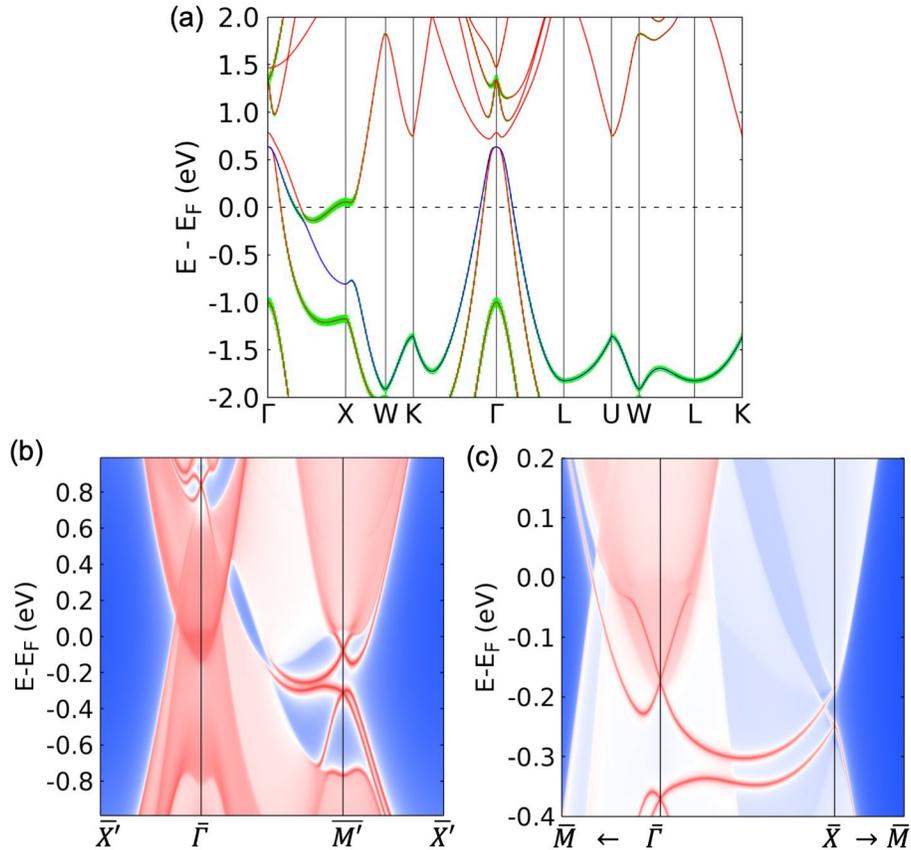

Figure S1. (a) Bulk band structure of non-magnetic (NM) NdBi. The top valence band is in blue and the green shade stands for the projection of Bi $p$ orbitals. (b) (001) surface spectral function of NM NdBi in the primitive tetragonal SBZ and (c) the folded tetragonal SBZ with zoom in energy range (see Fig.2(b) for the k-point labels).

# Supplementary Note 2: Multi-q magnetic space groups and bulk band structures

The three different multi-q AFM structures of NdBi are shown in Fig.1 (d), (g) and (j) with the corresponding bulk band structures presented in Fig.2 (a), (c) and (d), respectively. The 1q wave vector along the [001] direction (Fig.1 (d)) has a layered stacking AFM structure with Nd magnetic moments pointing along [001]. The associated tetragonal magnetic space group (MSG) $P_I4/mnc$ (or 128.410) has both inversion and non-symmorphic time-reversal symmetry (TRS) of {1'| ½ ½ ½}, where {1'} is TRS operation and {½ ½ ½} is half translation along all the three unit-cell vectors, respectively. Thus, all the bands are at least doubly degenerate in 1q. The bulk Brillouin zone (BZ) and high-symmetry points are shown in Fig.2(b) by the dashed lines. The NM NdBi is known to have a band inversion between the Bi $p$ and Nd $d$ orbitals with anti-crossing along the $\Gamma$-$X$ direction to host a strong topological insulator (TI) state (see Fig.S1(a)). As plotted in Fig.2(a), the 1q band structure along the $\Gamma$-$M'$ direction retains the same band inversion (neck-narrowing region) between the Bi $p$ and Nd $d$ orbitals. However, along the $\Gamma$-$Z$ direction with the AFM stacking along [001], the band-folding opens up a band gap with both the top valence and bottom conduction bands being derived from the Bi $p$ orbitals (green shade). The original band inversion is pushed up in energy and becomes upper valence bands. There is a band crossing along the $\Gamma$-$Z$ direction giving a pair of DPs protected by $C_4$ rotational symmetry with the momentum-energy of (0, 0, ±0.0603 1/Å, $E_F$+419.9 meV). With the inversion symmetry and parity eigenvalues at time-reversal invariant momentum (TRIM), $Z_4 = \left(\frac{1}{2}\right)\sum_{k \in TRIM}(n_k^+ - n_k^-) \ mod \ 4$ is 2, where $n_k^+$ and $n_k^-$ are the number of filled states with even and odd parity, respectively. Thus, NdBi 1q is a Dirac semimetal with $Z_4$=2.

The effects of the hybridization gap of Bi $p$ orbitals from the band-folding of the band-inversion region can also be clearly seen in 2q, which has two wave vectors, one along [100] and the other [010]. As shown in Fig.1(g), the Nd magnetic moments on the (001) plane point to the positive and negative [110] direction in alternating rows, and those on the neighboring (001) plane point to the [$\bar{1}$10] direction also in alternating rows. The 45-degree rotation doubles the unit cell on the (001) plane. The tetragonal MSG of 2q is $P_C4_2/nnm$ (or 134.481), which still has the inversion symmetry and non-symmorphic TRS from {1'| ½ ½ 0}, maintaining the band double degeneracy. The band-folding in 2q (Fig.2(c)) now also happens at the *X* point, half-way along the *Γ-M'* direction, which is highlighted by the green arrow in reference to the 1q band structure in Fig.2(a). Upon the band-folding in the *Γ-M'* direction, a bulk band gap of about 0.2 eV opens at the *X* point toward the *X-Γ* direction near $E_F$ (green horizontal arrow) due to the hybridization between the folded bands of the same Bi *p* orbitals. The original band inversion now becomes the upper valence bands and is pushed to higher energy. Interestingly, with the magnetic moments changing from [001] to in-plane directions in 2q, the hybridization gap along the *Γ-Z* direction is reduced from that in 1q and two new band crossings emerge. These band crossings form two pairs of DPs. With the inversion symmetry and parity eigenvalues at TRIM, the calculated $Z_4$ index is again 2. Thus, NdBi 2q is also a Dirac semimetal, but in contrast to 1q, the bulk hybridization gap appears in the $k_x$-$k_y$ plane instead of the $k_z$ direction.

Moving on to the 3q structure, the three wave vectors along the [100], [010] and [001] directions generate an alternating all-in-all-out magnetic configurations along the {111} directions shown in Fig.1(j). The MSG of 3q is *Pn-3m'* (or 224.113), which is cubic and still has the inversion symmetry. But non-symmorphic TRS is broken and the double degeneracy is lifted along most of the high-symmetry directions except for {001}. Comparing the band structure of 3q in Fig.2(d) to 2q in Fig.2(c), the *Γ-X* and *Γ-Z* directions now become equivalent and so are the hybridization gaps. With lifting of the band double degeneracy along the other directions, new band crossings emerge around the *Γ* point, giving Weyl points (WP) along the *Γ-R* direction. Because of the inversion symmetry, the $Z_4$ index can still be calculated and it equals 2. Thus, 3q is a Weyl semimetal with the hybridization band gap along all the {001} directions. Interestingly, $Z_4$=2 for all the three gapless phases in different multi-q structures, because the sum of the difference in the number of opposite parity eigenvalues over TRIM does not change.

## Supplementary Note 3: 1q surface band structures on other surfaces

Surface spectral functions of NdBi 1q on the tetragonal top (001) and (010) side surface (equivalent to cubic (110)) are presented in Fig.S2. Along the $\bar{\Gamma}$-$\overline{M'}$ direction, the band inversion in NM remains and the associated surface states are evident in Fig.S2(a). In contrast along the $\bar{\Gamma}$-$\bar{Z}$ direction in Fig.S2(c), there is a bulk hybridization gap opening due to the AFM band-folding along the $k_z$ direction as discussed in Fig.2(a) of the main text. Unlike the unconventional surface state pair (SSP) on the (110) side surface (equivalent to cubic (010)) in Fig.3, there is only one surface state in the hybridization gap. The corresponding 2D surface Fermi surface at $E_F$ are plotted in Fig.S2(b) and (d), which do not match the ARPES data below $T_N$.

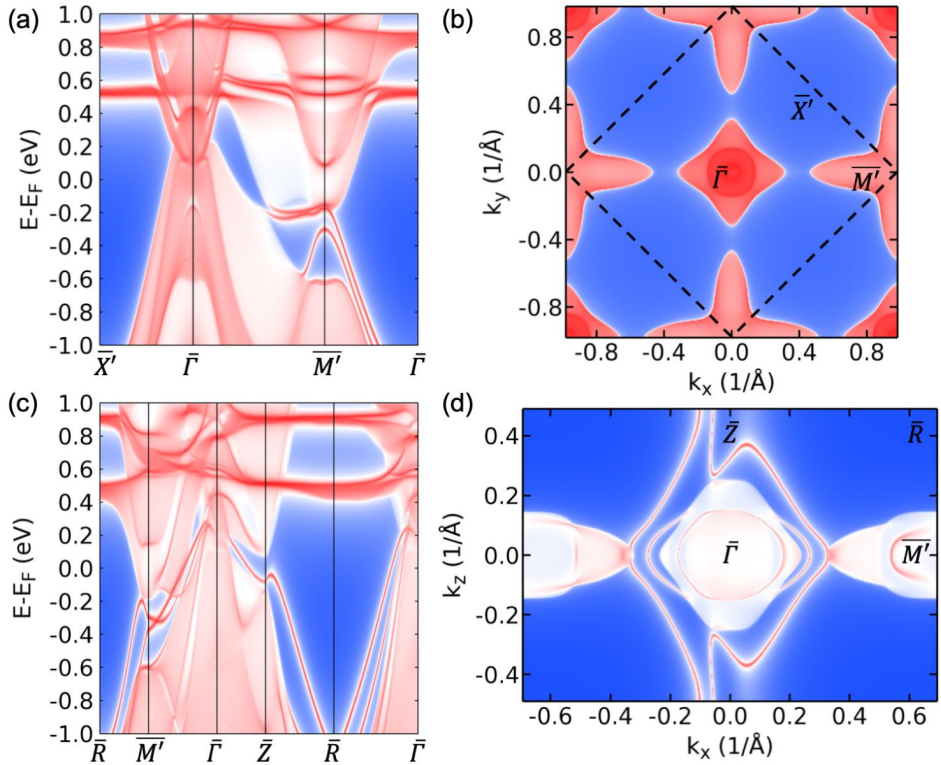

Figure S2. (a) Surface spectral function and (b) 2D Fermi surface at $E_F$ of NdBi 1q on tetragonal (001) surface (dashed square for the surface Brillouin zone). (c) and (d) Those on tetragonal (010) (equivalent to cubic (110)) surface.

# Supplementary Note 4: Comparison for 2-fold SSP of 2q and 1q to ARPES

Surface fermi surfaces (SFS) and spectral functions of NdBi 2q on (010) and 1q on the tetragonal (110) (or cubic (010)) side surfaces are compared with ARPES (Fig.S3) in the two sets of cuts, the same as those in Fig.4 for 2q (001). Although the SFS with the unconventional surface states pair is 2-fold in these two cases, the calculated spectral functions show a good agreement to ARPES on surface state dispersion across different cuts over one pair of the hole Fermi arc-like feature and surface state electron pocket, and also to 2q (001) with the 4-fold SSP in Fig.4. Thus, a mixture of {001} domains in 2q or 1q with different magnetic moment orientations can give an effective 4-fold SSP, especially putting the bulk band projection as background underneath the SSP, in a good agreement to ARPES.

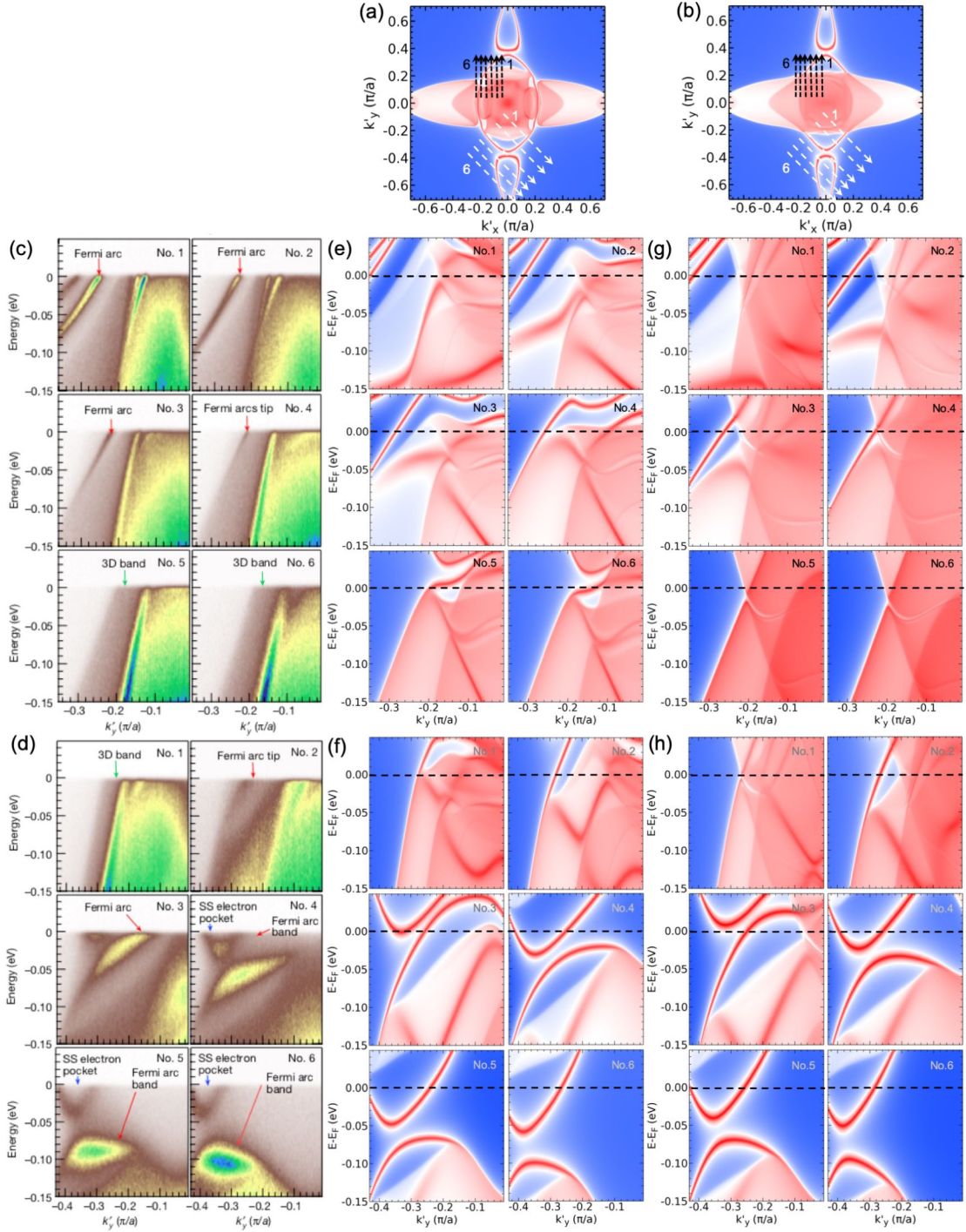

Figure S3. SFS on (a) 2q (010) at $E_F$–0.010 eV and (b) 1q (110) at $E_F$+0.030 eV labeled with two sets of cuts in black and white dashed lines. (c), (e) and (g) Comparison between ARPES and the calculated spectral functions at the cuts (black dashed lines in (a) and (b)) of 0.03, 0.08, 0.12, 0.14, 0.18 and 0.20 in the unit of the original NM SBZ. (d), (f) and (h) Comparison at cuts in the other direction (white dashed lines in (a) and (b)) at 0.03, 0.09, 0.17, 0.22, 0.24 and 0.27. Energies are with the shifts in (a) and (b), respectively.

## Supplementary Note 5: 3q (111) surface projection

For 3q Weyl semimetal with the 16 Weyl points in two sets (Fig.5(a)), we have also projected to the (111) surface, where only projections of 4 WPs in opposite chirality are overlapped at the $\bar{\Gamma}$ point, the rest 12 WPs are projected onto separate points in the (111) SBZ. As plotted in Fig.S4, these WP projections are connected by Fermi arcs spanning along the $\bar{M}$-$\bar{K}$-$\bar{M}$ direction.

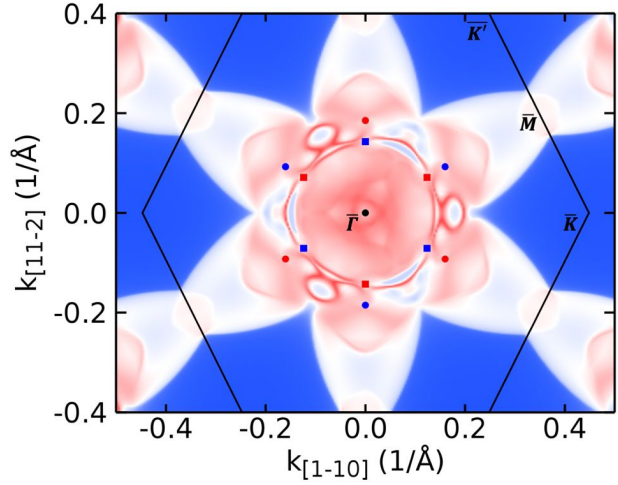

Figure S4. Surface Fermi surface near the Weyl point energy at $E_F+0.014$ eV on the 3q (111) surface. Projection of 12 WPs are labeled by the squares and circles in red (blue) for chirality +1(−1). The projection of 4 WPs of opposite chirality overlapped at the $\bar{\Gamma}$ point is labeled by a black circle.

## Supplementary Note 6: Dirac points in 2q

The 2q is in MSG of $P_C4_2/nnm$ (or 134.481) with both inversion and non-symmorphic TRS symmetry to have band double degeneracy. The top valence and bottom conduction bands cross twice along the $\Gamma$-$Z$ direction with different orbital content. We found two pairs of DPs as protected by the 4-fold screw axis (see Fig.S5).

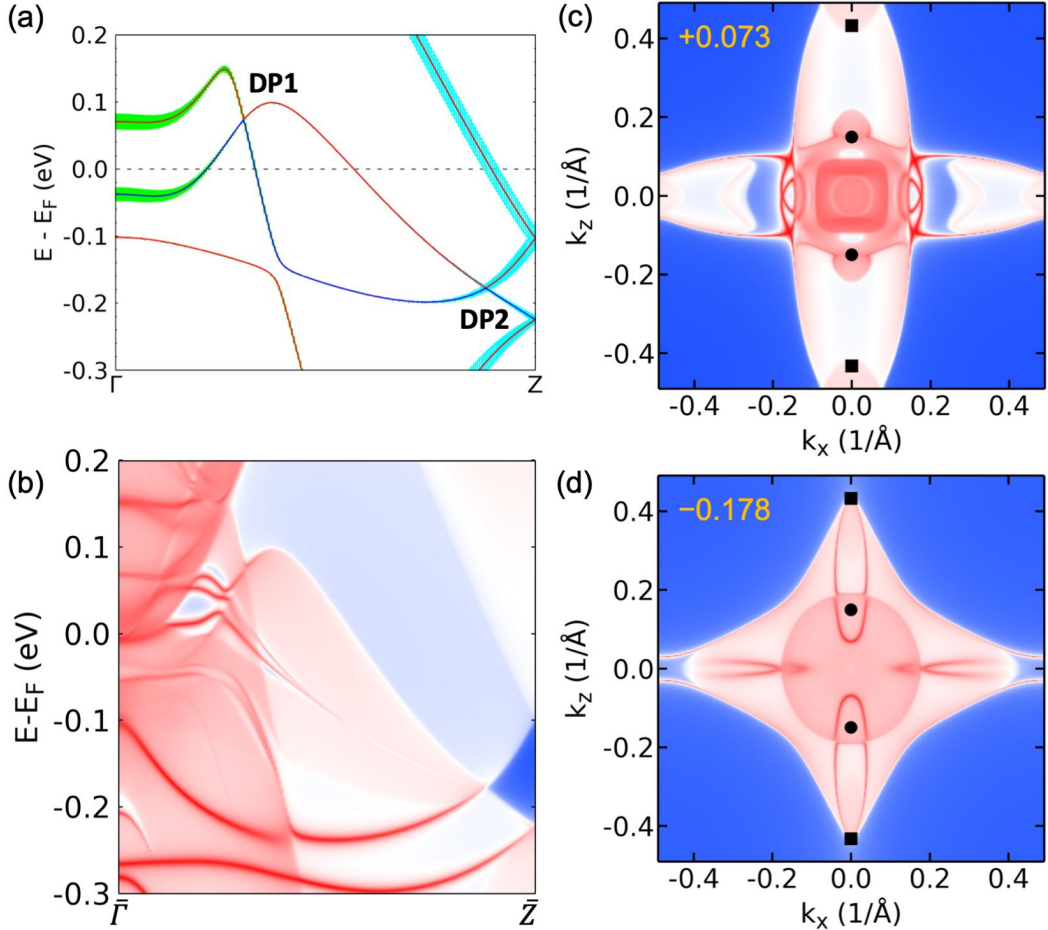

Figure S5. (a) Bulk band structure of NdBi 2q along the $\Gamma$-$Z$ direction. The top valence band is in blue and the green (cyan) shade stands for the projection of Bi $p_z$ (Nd $d_{xy}$) orbitals showing the two pairs of Dirac points (DPs) with the momentum-energy of DP1 (0, 0, ±0.1501 1/Å, $E_F$+73.0 meV) and DP2 (0, 0, ±0.4330 1/Å, $E_F$–177.5 meV), respectively. (b) Surface spectral function of 2q (010) along the $\bar{\Gamma}$-$\bar{Z}$ direction. Surface Fermi surface at the binding energies of the two DPs of (c) $E_F$+0.073 eV and (d) $E_F$–0.178 eV with the locations of the DP projections labeled by the black circles (DP1) and squares (DP2).

## Supplementary Note 7: DFT total energy with relaxation

The total energy differences among 1q, 2q and 3q with the fully relaxed unit cells calculated in PBEsol+U+SOC are plotted in Fig.S6(a). The energy difference between multi-q and 1q structures becomes smaller with higher U value, and 2q is more stable than 3q. With the U value larger than 9.6 eV, the 2q becomes the most favorable. The spin and orbital moments are around 3.0 and 5.9 $\mu_B$ in the opposite direction with high U values as Nd $4f$ moving away from the $E_F$. Thus, the net total magnetic moment on Nd is about 2.9 $\mu_B$. Bulk band structure of 2q with U=9.8 eV in Fig.S6(b) shows the same type of bulk hybridization gap opening similar to Fig.2(c). The surface spectral function in Fig.S6(c) and (d) on (001) and (010) show the unconventional SSP also emerge and have the same connectivity in the hybridization gap. But the splitting between the SSP is smaller than the case with U=6.3 eV and J=0.7 eV, because the Nd $4f$ bands are far away from the $E_F$.

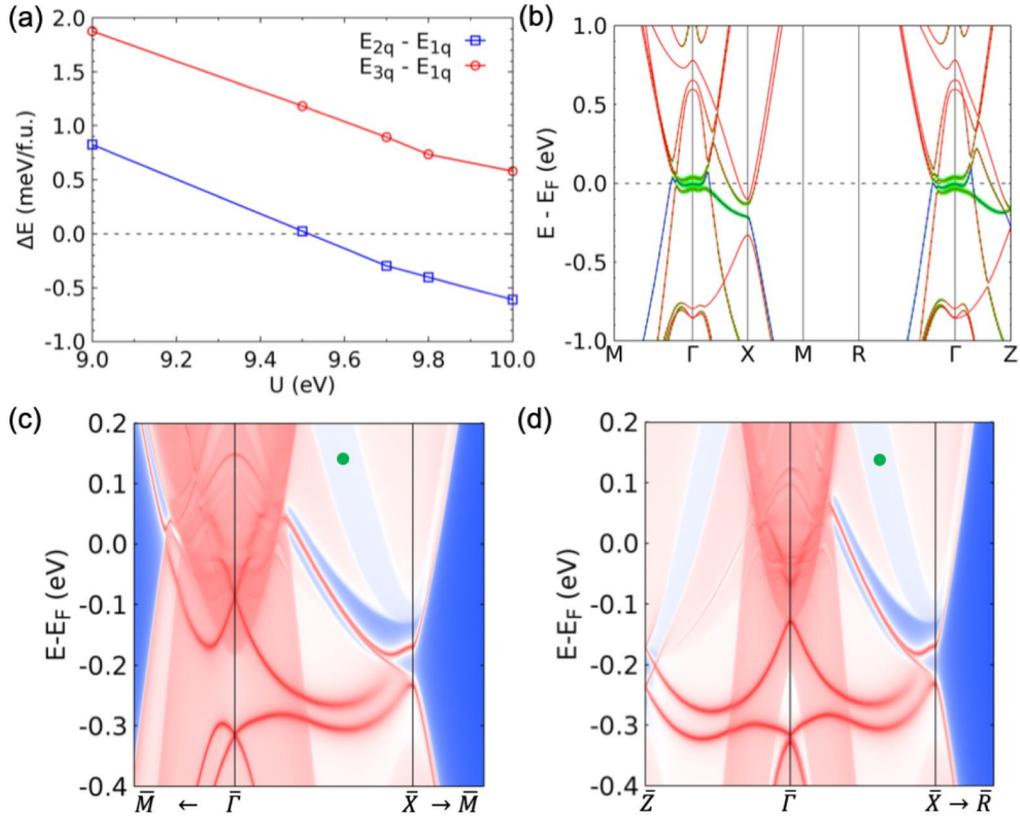

Figure S6. (a) DFT total energy difference as a function of U value among the 1q, 2q and 3q of the fully relaxed cell calculated with PBEsol+U+SOC. (b) Bulk band structure of 2q with U=9.8 eV. The top valence band is in blue and the green shade stands for the projection of Bi $p$ orbitals. (c) and (d) Surface spectral functions of NdBi 2q on (001) and (010), respectively.